\documentclass[conference]{IEEEtran}
\IEEEoverridecommandlockouts
\usepackage{cite}
\usepackage{amsmath,amssymb,amsfonts}
\usepackage{amsthm}
\usepackage{float}
\usepackage{graphicx}
\usepackage{textcomp}
\usepackage{xcolor}
\usepackage{booktabs}
\usepackage{multirow}
\usepackage{array}
\usepackage{enumitem}
\usepackage{url}
\usepackage{balance}
\usepackage{tikz}
\usepackage{pgfplots}
\pgfplotsset{compat=1.18}
\usepgfplotslibrary{statistics}
\usetikzlibrary{shapes.geometric,arrows.meta,positioning,fit,backgrounds,calc}

\newcommand{\dataset}{\textsc{OsintCI-66}}

\begin{document}

\title{A Case-Control Measurement Study of OSINT Source Effectiveness for Critical Infrastructure Defense}

\author{%
\IEEEauthorblockN{Ekrem E. Emeksiz\IEEEauthorrefmark{1},
Jeel Piyushkumar Khatiwala\IEEEauthorrefmark{1},
Divyangkumar Patel\IEEEauthorrefmark{2},
Weifeng Xu\IEEEauthorrefmark{1}}
\IEEEauthorblockA{\IEEEauthorrefmark{1}The University of Baltimore, Baltimore, MD, USA \quad
\IEEEauthorrefmark{2}Cox Automotive Inc, USA}
\IEEEauthorblockA{\{ekrem.emeksiz, jeel.khatiwala, wxu\}@ubalt.edu \quad divyangpatel38@gmail.com}
}

\maketitle

\begin{abstract}
Defenders of critical infrastructure (CI) subscribe to many public open-source
intelligence (OSINT) feeds without an empirical basis for which feeds actually
precede attacks. We provide one. Across 54 confirmed CI cyberattacks from 2010
through 2024 spanning twelve named CI sectors plus a cross-sector category
(consolidation rules in \S\ref{sec:audit}), paired with
12 null-control vulnerability
cases drawn from the same source space, we audit per-source attack coverage,
null-case contamination, and signal lead time for ten public OSINT source
classes that meet a minimum-volume threshold. Sources separate cleanly into
three operationally distinct mission profiles (pooled Fisher exact
$p = 3.4 \times 10^{-8}$): \textit{precursor} (six classes with zero observed
null firings at coverage at or above 5\%), \textit{disclosure-exposure} (three
classes whose null contamination meets or exceeds attack coverage), and one
large \textit{broad-coverage} class that mixes the two profiles but retains
91.3\% within-corpus precision. The precision-side classification is
stable across a 2019 temporal partition and across a US-versus-non-US
geographic partition.
Two sources, one broad-coverage and one precursor, cover 92.6\% of corpus
attacks; three cover 96.3\%.
The greedy portfolio at $k{=}3$ outperforms the mean
random three-source subset by 39.8 percentage points. Several source classes
widely treated as canonical for industrial control system defense fall into the
disclosure-exposure profile by operational mission, not by quality. Per-sector,
per-actor, and per-jurisdiction portfolios diverge in rank order despite a
shared rank-one source. The corpus, linkage protocol, and classification
rules are released.
\end{abstract}

\begin{IEEEkeywords}
Open-source intelligence (OSINT), cyber threat intelligence (CTI), threat
intelligence feeds, critical infrastructure security, ICS/SCADA, source
selection, attack precursors, empirical security measurement, case-control
study, set cover.
\end{IEEEkeywords}

\section{Introduction}
\label{sec:intro}

%
Critical-infrastructure (CI) defenders rely on numerous public
open-source intelligence (OSINT) sources: government advisories,
vulnerability databases, vendor threat reports, malware repositories, and
threat-exchange platforms; the CTI literature supplies rich
post-collection
frameworks~\cite{hutchins2011killchain,caltagirone2013diamond,strom2018attack}.
Security teams, however, have limited monitoring and triage capacity: a
SOC at a mid-sized utility cannot continuously review, validate,
correlate, and operationalize every available source, forcing a priority
decision among them. Yet there is little empirical evidence
showing which public sources are most strongly associated with confirmed
attacks rather than with general vulnerability disclosure or exposure
information; industry reputation tracks institutional authority, vendor
marketing, and historical prominence, none of which necessarily reflects
attack-precursor effectiveness. This paper addresses the problem of
evaluating public OSINT source effectiveness and of selecting a limited
portfolio of sources that maximizes coverage of confirmed CI
cyberattacks while accounting for null-case contamination, signal
redundancy, and source mission.

We provide that empirical evidence. Across a fifteen-calendar-year corpus
(2010--2024) of 54 confirmed CI cyberattacks paired with 12 null-control
vulnerability cases drawn from the same OSINT source space, we audit which
sources fire on which kinds of events, with what lead time, and with what
contamination. The methodology adapts the case-control study design Allodi
and Massacci~\cite{allodi2014comparing} introduced from epidemiology to
vulnerability exploitation analysis, applied here to OSINT source
evaluation. The work sits at the intersection of network-security
telemetry, threat-intelligence sharing, and operational SOC design that
defines the IEEE CNS scope.

\subsection{Contributions}

\textbf{C1: Case-control methodology for OSINT source evaluation.} We bring
paired attack-and-null measurement to CTI source assessment. Prior
work~\cite{li2019tealeaves,bouwman2020tivalue} measures source behavior on
attack-related ground truth alone. Pairing attack incidents with null-control
cases drawn from the same source space lets us compute source precision
rather than coverage alone. To our knowledge this is the first application of
case-control study design~\cite{allodi2014comparing} to OSINT \emph{source}
evaluation rather than vulnerability exploitation prediction.
The null side is modest at $n{=}12$, a constraint on interval width rather
than on the structural finding.

\textbf{C2: An empirically derived, stable source-mission taxonomy.} OSINT
source classes separate into three operational profiles by joint
attack-coverage and null-contamination behavior (pooled Fisher exact
$p = 3.4 \times 10^{-8}$). Six classes display the \emph{precursor} pattern
(zero observed null firings at coverage between 5.6\% and 27.8\%);
the 12-null sample leaves individual
precursor classes under-powered after Benjamini-Hochberg correction (the
Clopper-Pearson upper bound at zero firings is 22.1\%), so the pattern is
reported at the family level (\S\ref{sec:taxonomy}).
Three classes display
\emph{disclosure-exposure} behavior with within-corpus precision between
33.3\% and 69.2\% paired with null contamination at or above attack
coverage. One class, vendor-tier research, occupies a \emph{broad-coverage}
position (77.8\% coverage, 33.3\% null contamination, 91.3\% precision).
The precision-side classification is stable across a 2019 temporal
partition and across a US-versus-non-US geographic partition. Several
classes commonly treated as ICS-defender canon fall in the
disclosure-exposure profile by operational mission rather than quality.

\textbf{C3: Constrained set-cover formulation of CTI source-portfolio
selection.} We formulate source selection as cardinality-constrained maximum
coverage and solve it with greedy approximation under the $(1{-}1/e)$
guarantee.
A two-source portfolio, one broad-coverage plus one precursor class, covers
92.6\% of corpus attacks (95\% bootstrap CI 85.2--98.1\%); three sources
cover 96.3\% (92.6--100.0\%).
The greedy
portfolio at $k{=}3$ outperforms the mean random three-source subset by 39.8
percentage points, quantifying the value of source-aware portfolio
construction.

\textbf{C4: A released null-controlled OSINT corpus.} \dataset{} (15 years, 54
attacks, 12 nulls, 161 verified signals across 15 observed source classes)
was constructed under a two-verifier-with-adjudication protocol.
The
protocol and a post-build release-preparation pass are documented in
\S\ref{sec:audit}; pre-adjudication agreement was 84.5\% across 251
protocol outcomes.
To our knowledge this is the first publicly
released null-controlled OSINT corpus for CI security research with a
documented adjudication protocol and full audit trail.

\textbf{Roadmap.} \S\ref{sec:problem} fixes the defender model, metrics,
and set-cover objective; \S\ref{sec:audit} constructs \dataset{}; \S\ref{sec:taxonomy} derives the three-profile taxonomy
from joint coverage and contamination behavior; \S\ref{sec:predval} tests
its stability under temporal and geographic partition;
\S\ref{sec:portfolio} through \S\ref{sec:walked} translate the
measurements into portfolios, suppression bounds, and lead times. A
signal is \emph{pre-attack} before the verified attack date,
\emph{same-day} at day zero, and \emph{post-attack} when it accompanies
incident response; coverage counts all three, lead-time analysis only
the first two.

\section{Background and Related Work}
\label{sec:related}

\textbf{Comparative CTI source measurement.} Li et
al.~\cite{li2019tealeaves} measure precision-recall variance across CTI
feed types; Bouwman et al.~\cite{bouwman2020tivalue} find 2.5--4.0\%
vendor-to-vendor overlap, with follow-on
work~\cite{bouwman2022threats,wagner2023sharing} confirming structural
heterogeneity; TTP and quality
mining~\cite{husari2017ttpdrill,kuehn2024bandit,sun2023ctimining,schaberreiter2019quality}
treats OSINT as content substrate. None addresses source-level attack
precedence against null controls.
On the forensic side, hypothesis-driven search recovers identifiers
(emails, phone numbers, domains) from mobile application databases at
scale~\cite{khatiwala2026pii}, the pivot artifacts the
infrastructure-linkage criterion of \S\ref{subsec:linkage} consumes, and
case-study evidence documents reliability limits of LLM-surfaced forensic
evidence~\cite{khatiwala2025reliability}; both motivate measuring the
source layer itself.
A reproducible benchmark for structural inference over undocumented mobile
databases extends the same evaluation discipline to agentic
reasoning~\cite{khatiwala2026structural}.

\textbf{Vulnerability exploitation prediction and CI security.} Allodi and
Massacci~\cite{allodi2014comparing} brought case-control design to
security research. Twitter-based prediction~\cite{sabottke2015twitter},
exploit-timing analysis~\cite{householder2020exploit}, and
EPSS~\cite{jacobs2021epss} operate per-CVE; we operate per-source and
treat EPSS as complementary. NIST SP 800-82r3~\cite{stouffer2023nist}
fixes the defender model; Dragos~\cite{dragos2024report} is itself
evaluated. Set cover has prior IDS and sensor
use~\cite{noel2008ids,krause2008sensor} under the classical
guarantee~\cite{nemhauser1978submodular}; empirical case-control priors
for CTI source portfolios are new.

\section{Problem Formulation}
\label{sec:problem}

A CI defender selects a portfolio $\mathcal{P} \subseteq \mathcal{S}$ from the
universe $\mathcal{S}$ of available public sources, with the aim of maximizing
the fraction of future attacks covered by at least one signal from
$\mathcal{P}$, subject to a monitoring budget. We define four metrics per
source $s$. A \emph{pre-attack signal} is a publicly observable data point
published before a confirmed attack with verifiable linkage via CVE, malware
family, threat actor, or infrastructure (\S\ref{subsec:linkage}).
\emph{Source attack coverage} over the attack corpus $\mathcal{A}$ and
\emph{source precision} are
\begin{align}
\mathrm{cov}(s) &= \bigl|\{\, i \in \mathcal{A} : s \in \mathcal{S}_i \,\}\bigr| \;/\; |\mathcal{A}|,\\
\mathrm{prec}(s) &= \frac{\mathrm{cov}(s)\,|\mathcal{A}|}{\mathrm{cov}(s)\,|\mathcal{A}| + \mathrm{null}(s)\,|\mathcal{N}|},
\end{align}
where \emph{null contamination} $\mathrm{null}(s)$ is the coverage
analogue over the null corpus $\mathcal{N}$, and $\mathrm{prec}(s)$ is the
within-corpus fraction of incidents fired on by $s$ that are confirmed
attacks.
\footnote{Within-corpus precision
is the relevant quantity for source \emph{ranking}. It is not a
deployment-time positive predictive value because deployment base rates
differ from the corpus 54:12 ratio.}

\textbf{Coverage scope.} An incident counts as covered by source $s$ when
$s$ produced \emph{any} verified-linkage signal for that incident,
irrespective of whether the signal preceded the attack or accompanied
incident response. The set $\mathcal{A}_s$ is therefore the set of attacks
on which $s$ eventually fired.
Pre-attack lead time is reported separately
(\S\ref{sec:walked}) over signals with days-before-attack $\geq 0$; a
strict pre-attack-only recompute of every per-source result
(\S\ref{sec:limits}) preserves the taxonomy and the rank-one finding.

\textbf{Threat model.} We model a SOC, threat-intelligence team, or CISO
office choosing OSINT sources under finite attention, facing ransomware
operators, nation-state actors, and hacktivists, consistent with NIST SP
800-82r3~\cite{stouffer2023nist}; adaptive signal suppression is treated
structurally in \S\ref{sec:adversary}.

\textbf{Research questions.} (RQ1) Do public OSINT sources separate into
mission profiles by their joint attack-coverage and null-contamination
behavior, and are these profiles statistically distinguishable, operationally
interpretable, and stable under temporal and geographic partition? (RQ2)
Given attention budget $k$, which $k$ sources cover the largest fraction of
confirmed CI attacks, and how quickly does marginal coverage decline? (RQ3)
Do optimal portfolios differ by sector, actor type, and jurisdiction?

\textbf{Set-cover formulation.} Let $x_s \in \{0,1\}$ indicate whether source
$s$ is monitored and $y_i \in \{0,1\}$ whether attack $i$ is reached by at
least one selected source. The cardinality-constrained maximum-coverage
problem is
\begin{align}
\max\ & f(\mathcal{P}) := \textstyle\sum_{i \in \mathcal{A}} y_i \\
\mathrm{s.t.}\ & y_i \leq \textstyle\sum_{s \in \mathcal{S}_i} x_s\ \forall i,\
                \textstyle\sum_{s} x_s \leq k,\ x_s, y_i \in \{0,1\}. \nonumber
\end{align}
The coverage function $f(\mathcal{P}) = |\bigcup_{s \in \mathcal{P}}
\mathcal{A}_s|$, where $\mathcal{A}_s = \{i : s \in \mathcal{S}_i\}$, is
monotone submodular. Greedy optimization achieves $(1{-}1/e) \approx 63\%$ of
the optimum under cardinality constraint~\cite{nemhauser1978submodular}, and
we verify against exhaustive enumeration in \S\ref{sec:portfolio}. The
empirical priors $\mathrm{cov}(s)$ and $\mathrm{prec}(s)$ that drive the
optimization are this paper's primary measurement contribution.

\section{Empirical Audit}
\label{sec:audit}

\begin{table*}[t]
\caption{Per-Source Attack Coverage, Null Contamination, and Mission
Profile Across 54 Confirmed CI Cyberattacks and 12 Null-Control Cases,
2010--2024.}
\label{tab:sourcetable}
\centering
\footnotesize
\setlength{\tabcolsep}{8pt}
\renewcommand{\arraystretch}{1.05}
\begin{tabular}{@{}lrrrrrrrl@{}}
\toprule
\textbf{Source class} & \textbf{Total} & \textbf{Att.\ inc.} &
\textbf{Null inc.} & \textbf{Att.\ cov.} & \textbf{Null cont.} &
\textbf{Prec.\ (95\% LB)} & \textbf{Med.\ lead} & \textbf{Profile}\\
& \textbf{signals} & & & (\% of 54) & (\% of 12) & (\%) & \textbf{(days)} & \\
\midrule
\multicolumn{9}{l}{\textit{Broad-coverage}}\\
Vendor-Tier1                   & 63 & 42 & 4 & 77.8 & 33.3 & 91.3 (83.1) & 24 & Broad-coverage\\
\midrule
\multicolumn{9}{l}{\textit{Precursor}}\\
US-CERT (TA/AA alerts)         & 15 & 15 & 0 & 27.8 &  0.0 & 100.0 (78.5) & 32 & Precursor\\
VirusTotal                     &  7 &  6 & 0 & 11.1 &  0.0 & 100.0 (50.2) &  0 & Precursor\\
NCSC-Tier1 (UK/DE/FR/CA)       &  6 &  6 & 0 & 11.1 &  0.0 & 100.0 (50.2) & 13 & Precursor\\
MalwareBazaar                  &  5 &  5 & 0 &  9.3 &  0.0 & 100.0 (43.1) & 11 & Precursor\\
CERT-UA                        &  4 &  4 & 0 &  7.4 &  0.0 & 100.0 (34.4) & 24 & Precursor\\
OTX (open threat exchange)     &  3 &  3 & 0 &  5.6 &  0.0 & 100.0 (23.8) & 17 & Precursor\\
\midrule
\multicolumn{9}{l}{\textit{Disclosure-exposure}}\\
CISA-KEV                       & 16 &  9 & 4 & 16.7 & 33.3 & 69.2 (39.9) & 24 & Disclosure-exp.\\
NVD/NIST                       & 24 & 11 & 11 & 20.4 & 91.7 & 50.0 (34.9) & 62 & Disclosure-exp.\\
ICS-CERT (ICSA)                & 12 &  4 & 8 &  7.4 & 66.7 & 33.3 (11.6) & 420 & Disclosure-exp.\\
\midrule
\multicolumn{9}{l}{\textit{Below-threshold ($<$3 signals each, in released corpus): Sector-ISAC, Dark-Web, GitHub-PoC, ExploitDB, Internal/HSE-IR}}\\
\bottomrule
\end{tabular}\\[2pt]
{\scriptsize Precision $=\mathrm{cov}(s)|\mathcal{A}|/[\mathrm{cov}(s)|\mathcal{A}| + \mathrm{null}(s)|\mathcal{N}|]$ which reduces to $a/(a+n)$ on attack-firing and null-firing counts, with $|\mathcal{A}|=54$, $|\mathcal{N}|=12$, conditional on the linkage protocol of \S\ref{subsec:linkage}. The 95\% lower bound on precision uses a joint Clopper-Pearson construction (upper bound on null contamination, lower bound on attack coverage). Median lead is the proper sample median over the set $\{\mathrm{dba}\geq 0\}$ rounded to the nearest day. The CISA-KEV class as recorded in this study aggregates the Known Exploited Vulnerabilities catalog~\cite{cisakev} with CISA-issued Joint Advisories; the borderline disclosure-exposure profile of the class reflects this aggregation.}
\end{table*}

\subsection{Corpus and methodology}

\textbf{Data processing pipeline.} Corpus production and analysis proceed
in nine stages: case selection $\to$ signal collection $\to$ linkage
$\to$ dual verification $\to$ cleaning $\to$ source aggregation $\to$
metric computation $\to$ statistical analysis $\to$ portfolio
optimization (\S\ref{sec:audit}--\S\ref{sec:portfolio}; metrics
\S\ref{sec:problem}).

\dataset{} comprises 54 confirmed CI cyberattacks and 12 null-control cases
from May 2010 through October 2024 across twelve named CI sectors plus a cross-sector category.\footnote{The released corpus CSV uses the disaggregated CISA sector names; the analysis consolidates pairs the CISA taxonomy treats as a single sector (Healthcare and Public Health into Healthcare; Critical Manufacturing into Manufacturing; Transportation Systems into Transport; Water and Wastewater into Water; Government Facilities into Government; Financial Services into Financial) before stratification. Actor-type labels Cybercrime and Cybercriminal are consolidated as Cybercriminal; Nation-State and Nation-State-Aligned as Nation-State. The corpus is referenced by the compact name \dataset{} in the paper; the released CSV filenames use the versioned tag v3.1.} The corpus distributes among Energy (15 attacks plus 4 nulls), Healthcare
(14 attacks plus 1 null), Manufacturing (5 plus 3), Transport (3),
Water (3), Government (3 plus 1), Financial (2), Food (2), Information
Technology (2), Commercial (2), Telecommunications (1), Supply Chain (1),
and Cross-sector (1 plus 3). Attack cases satisfy three criteria: (1)
confirmed operational impact verified through a Tier-1 source (CISA advisory,
SEC 8-K filing, government press release, or Congressional testimony); (2)
at least one OSINT signal qualifying under the linkage protocol of
\S\ref{subsec:linkage} and dated at or before the verified attack date;
(3) attack date verifiable within seven days. 
Criterion (2) admits same-day
qualifying signals so that near-zero-day events (such as MOVEit, where
CVE-2023-34362 mass exploitation began May 27, 2023 and first public
disclosure followed on May 31; \S\ref{sec:walked}) remain in scope as
operationally-near-zero-day limit cases rather than being excluded.
Null cases are vulnerabilities or sector-advisory events meeting CISA KEV
or ICS-CERT criteria, gated on CVSS $\geq 7.0$ or a qualifying ICS-CERT
advisory, deployed in at least one CI sector, for which no public reporting
of operational impact appeared within 180 days of the primary signal
date.%
\footnote{The 180-day null window exceeds the public-exploitation
timeline distribution reported by Householder et
al.~\cite{householder2020exploit} for the majority of CVEs.
Ten of the 12 null cases carry CVSS $\geq 7.0$; NC-010 qualifies through an
ICS medical-device advisory (ICSMA-23-194-01) at CVSS 6.6, and NC-011 is a
sector-bulletin archetype with no single CVE. All 12 sit in deployed CI
environments, the population most likely to generate detectable public
reporting if exploited.
Null
contamination by undetected attacks would \emph{attenuate} the observed
attack-versus-null separation, so the reported $p$-value is conservative
against this bias.
Absence of public impact reporting is an imperfect negative label; the
attenuation argument makes the reported separation a lower bound.
The null selection log is included in the released
corpus.}

\textbf{Signal-to-incident linkage protocol.}
\label{subsec:linkage}
A signal qualifies as linked when it satisfies at least one of seven
criteria. The \emph{strict} four (used for headline results) are: (a) CVE
match between signal and attack attribution; (b) malware-family match to
post-incident forensic attribution; (c) threat-actor match to publicly
attributed actor (the \texttt{actor} and \texttt{threat-actor} dataset
labels are equivalent and consolidate to this criterion); (d)
infrastructure match (IOC, IP, domain, or hash) to post-incident
forensics. The \emph{boundary} three (retained for completeness, annotated
in the corpus) are: (e) named-advisory cross-reference
(\texttt{advisory}/\texttt{advisory-echo}); (f) TTP match where
post-incident MITRE ATT\&CK characterization names a specific
technique-and-software combination present in the pre-attack signal
(\texttt{TTP}); (g) sectoral or geopolitical context match treated
separately in stratified analysis
(\texttt{sectoral}/\texttt{geopolitical}). Topical proximity alone does
not qualify. Post-attack incident-response advisories contribute to
coverage but are excluded from the lead-time analysis.
Of the 161 verified signals, 129 link under (a)--(d) (strict) and 32 under
(e)--(g) (boundary); a strict-only recompute preserves the precision-side
taxonomy and the rank-one finding (\S\ref{sec:limits}).

\textbf{Sources and aggregation.} We collected signals from public sources
across government Tier-1 advisory channels (US-CERT TA/AA alerts, ICS-CERT
ICSA bulletins, CISA KEV listings, allied national CSIRT advisories from the
NCSC-UK, BSI, ANSSI, and CCCS tier, CERT-UA), the canonical vulnerability
database (NVD), industry Tier-1 vendor research, public malware repositories
(MalwareBazaar, VirusTotal), public exploit code repositories (ExploitDB,
GitHub), sector information-sharing organizations, public threat-exchange
platforms (OTX), and dark-web monitoring.
After verification, 161 signal rows distribute across 15 observed
classes; 10 meet the 3-signal minimum and form the per-class table. The
threshold marks where Clopper-Pearson precision lower bounds stop being
informative (at $a{=}n{=}1$ the 95\% LB admits the full unit interval);
the 5 below-threshold classes remain in the release but are excluded from
recommendations. Aggregation rules appear in Table~\ref{tab:sourcetable}
and \S\ref{sec:taxonomy}.

\textbf{Verification protocol.} Two-verifier with adjudication: matched
independent reviews enter clean-pass or clean-fail; disagreements enter a
hard-conflict bucket resolved on evidence strength, or mixed-outcome for
full-team discussion. Across 251 outcome rows (66 incidents, 185
signals): 195 clean-pass, 17 clean-fail, 21 hard-conflict (adjudicated),
9 soft-conflict, 9 mixed-outcome; pre-adjudication agreement
84.5\%~\cite{landis1977kappa}, conflict envelope 12.0\%. Release
preparation surfaced 51 rows with failed Internet Archive captures; a
single-reviewer round 2 confirmed 21 URLs, replaced 24 with canonical
equivalents, retried 2, and removed 4.

\textbf{AI assistance.} AI assistance was used in select corpus-construction
support phases (URL substantiation, candidate-pool surfacing, sanity-check
scripts), reviewed for correctness, and not relied upon for final
row-level verification.
This division of labor reflects documented reliability limits of
LLM-discovered forensic evidence~\cite{khatiwala2025reliability}.

\subsection{Per-source measurement}

Table~\ref{tab:sourcetable} reports per-class verified signal counts, attack
incidents covered, null cases covered, source precision with Clopper-Pearson
95\% lower bounds, median attack-signal lead time, and the mission profile
assigned in \S\ref{sec:taxonomy}.

\textbf{Interpretation of precision values.} Precision in
Table~\ref{tab:sourcetable} measures attack-versus-null
distinguishability under \S\ref{subsec:linkage}. With zero null firings
over 12 nulls, the one-sided 95\% Clopper-Pearson bound on true
contamination is 22.1\%, so zero-null classes keep point precision 100\%
with lower bounds from 23.8\% to 78.5\% by coverage; we treat the lower
bound as the honest summary.


\section{Source Mission Taxonomy}
\label{sec:taxonomy}

The per-source measurements separate cleanly along the precision axis into
three operational profiles.

\begin{figure}[!b]
\centering
\begin{tikzpicture}
\begin{axis}[
  width=8.4cm, height=4.8cm,
  xlabel={Null contamination (\% of 12 nulls)},
  ylabel={Attack coverage (\% of 54 attacks)},
  xmin=-5, xmax=105, ymin=-3, ymax=85,
  grid=major, grid style={dashed,gray!30},
  legend pos=north east, legend style={font=\scriptsize},
  tick label style={font=\small},
  xlabel style={font=\small}, ylabel style={font=\small},
]
\addplot[only marks, mark=diamond*, mark size=4.5pt, color=violet]
  coordinates { (33.3, 77.8) };
\addlegendentry{Broad-coverage (1)}
\addplot[only marks, mark=*, mark size=3pt, color=blue]
  coordinates { (0, 27.8) (0, 11.1) (0, 11.1) (0, 9.3) (0, 7.4) (0, 5.6) };
\addlegendentry{Precursor (6)}
\addplot[only marks, mark=square*, mark size=3pt, color=red]
  coordinates { (33.3, 16.7) (91.7, 20.4) (66.7, 7.4) };
\addlegendentry{Discl-Exposure (3)}
\draw[dashed, gray!40, thin] (axis cs:5, 5) -- (axis cs:80, 80);
\node[font=\tiny, anchor=west, gray!55, rotate=33, fill=white, inner sep=0.5pt] at (axis cs:55, 60) {nct = cov};
\node[font=\scriptsize, anchor=west, color=violet!80!black] at (axis cs:36, 77.8) {Vendor-Tier1};
\node[font=\scriptsize, anchor=west] at (axis cs:3, 28) {US-CERT};
\node[font=\scriptsize, anchor=west, fill=white, inner sep=1pt] at (axis cs:3, 18.5) {VirusTotal, NCSC};
\node[font=\scriptsize, anchor=west, fill=white, inner sep=1pt] at (axis cs:3, 10) {MalwareBazaar};
\node[font=\scriptsize, anchor=west, fill=white, inner sep=1pt] at (axis cs:3, 1.5) {CERT-UA, OTX};
\node[font=\scriptsize, anchor=west, color=red!80] at (axis cs:36, 16.7) {CISA-KEV};
\node[font=\scriptsize, anchor=east, color=red!80] at (axis cs:89, 20.4) {NVD};
\node[font=\scriptsize, anchor=west, color=red!80] at (axis cs:69, 7.4) {ICS-CERT};
\end{axis}
\end{tikzpicture}
\caption{Empirical mission classification on the attack-coverage versus
null-contamination plane. The broad-coverage class (Vendor-Tier1) sits to
the upper-left of the diagonal $\mathrm{nct}{=}\mathrm{cov}$ guide; the
disclosure-exposure cluster sits to the lower-right of the same diagonal.
The diagonal corresponds to a base-rate-balanced precision of 50\%.}
\label{fig:scatter}
\end{figure}
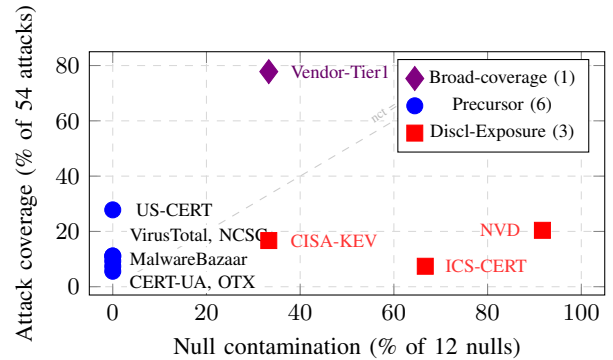

\textbf{Precursor sources} fire when attacks are imminent or underway,
publishing indicators tied to active threat actors, fresh malware samples,
observed exploitation, or attribution specific to ongoing campaigns.
Six classes display
this behavior with zero observed null
firings and attack coverage between 5.6\% and 27.8\%. US-CERT TA/AA alerts
lead this group at 27.8\% coverage with 32-day median lead. The malware
repositories MalwareBazaar (9.3\%, 11-day lead) and VirusTotal (11.1\%, same-day)
exemplify the tight-coupling case where sample first-submission tracks
attack onset closely.
CERT-UA (7.4\%, 24-day median) reflects one operational window: all four
signals fall on February 23--25, 2022, immediately before the Russian
invasion of Ukraine, covering two Sandworm-attributed incidents (Ukrenergo
INC-010, Viasat INC-011), Costa Rica Conti INC-029 via a Conti leak 51
days ahead, and Kojima INC-036 under \texttt{geopolitical} boundary
linkage. The 24-day median is the midpoint of a bimodal set (1, 3, 44,
51); the class's general behavior is closer to concurrent reporting, and
the v3.1 lead profile is specific to the pre-invasion window.

\textbf{Disclosure-exposure sources} fire on disclosure or observed
attack surface regardless of whether an attack follows; they serve
patch-management functions, and treating them as precursors generates
non-actionable incident-response alerts. Three classes show the signature
where null contamination meets or exceeds attack coverage.
NVD (precision 50.0\%,
null contamination 91.7\%) is the canonical vulnerability database and
fires on nearly every null case by construction. ICS-CERT (33.3\%, 66.7\%)
fires on a majority of the ICS null cases. CISA-KEV (69.2\%, 33.3\%) sits
at the boundary; it adds CVEs on exploited-in-wild evidence but also lists
high-severity CVEs that were never operationally confirmed against the 12
null cases. The class label aggregates the KEV catalog with the CISA Joint
Advisories appearing in the corpus
(AA21-131A~\cite{cisaaa21131a}, AA23-144A~\cite{cisaaa23144a},
AA23-158A~\cite{cisaaa23158a}, AA23-353A~\cite{cisaaa23353a},
AA24-038A~\cite{cisaaa24038a}); the released corpus disambiguates the two
per signal.


\textbf{Broad-coverage} is the empirical position occupied by the
\emph{Vendor-Tier1} class: 42 attack incidents covered against 4 null
firings (77.8\% coverage, 33.3\% null contamination, 91.3\% within-corpus
precision, 95\% CP lower bound 83.1\%). The class aggregates threat
research from Tier-1 industry vendors and major incident-response firms;
per-vendor decomposition would split a subset into pure-precursor and a
different subset into disclosure-exposure.
We retain the aggregation as the most useful operational description: it
anchors most portfolios while requiring downstream filtering of the
disclosure-style content it also carries.
Vendor-Tier1 is a measurement category, not a subscribable product;
guidance phrased on it means monitoring the tier's public research
output.

\textbf{Statistical validation.} Pooled across the 6 precursor classes,
attack-incident firings number 39 against 0 null-incident firings; pooled
across the 3 disclosure-exposure classes, 24 against 23. The
contingency $\bigl[\begin{smallmatrix}39 & 0\\ 24 & 23\end{smallmatrix}\bigr]$
yields one-sided Fisher exact $p = 3.4 \times 10^{-8}$ (two-sided
$3.9 \times 10^{-8}$) against equal attack-to-null firing ratio. The
pooled separation is highly significant on the small-$n$ corpus.

The pooled contingency counts class-incident firings, so one incident
covered by several classes contributes several counts. Collapsing to
incident granularity removes the dependence: at least one precursor class
fires on 31 of 54 attacks and 0 of 12 nulls (one-sided Fisher exact
$p = 1.7 \times 10^{-4}$, equal to the exact probability that a random
relabeling of the 66 incidents places no null among the 31 precursor-fired
incidents), and at least one disclosure-exposure class fires on 21 of 54
attacks and 12 of 12 nulls (two-sided $p = 1.4 \times 10^{-4}$); a label
permutation of the pooled statistic finds none as extreme in
$2 \times 10^{5}$ draws ($p < 5 \times 10^{-6}$). The pooled
$3.4 \times 10^{-8}$ is a firing-level figure.

To address the concern that pooling may mask within-class heterogeneity,
we run per-source Fisher exact tests against the corpus baseline attack
proportion (54 of 66 incidents, 81.8\%) for the 11 classes with at least
two combined firings, applying Benjamini-Hochberg correction at FDR 0.05.
Three classes survive: Vendor-Tier1 ($q^* = 0.0174$), NVD ($q^* = 0.0001$),
and ICS-CERT ($q^* = 0.0002$).
Remaining classes lack 12-null-side power even where point behavior is
clean; the broad-coverage class and the two largest disclosure-exposure
classes are individually distinguishable, the pooled separation is highly
significant, and individual precursor significance awaits a larger null
corpus.

Figure~\ref{fig:scatter} places the three profiles in distinct regions of
the coverage-contamination plane, split by the
$\mathrm{nct}{=}\mathrm{cov}$ diagonal.


\section{Stability Under Temporal and Geographic Partition}
\label{sec:predval}

A taxonomy derived from the same data on which it is evaluated can be
descriptive without being predictive. We address this with two out-of-sample
partitions.

\textbf{Temporal partition.} We split attacks at calendar year 2019 (11
pre-2019, 43 in 2019 through 2024). The pre-2019 partition is small but
admits the test on at least the rank-one source. On both subsets the
broad-coverage class Vendor-Tier1 retains rank-one greedy position and
the precision-side classification of the disclosure-exposure classes (NVD,
ICS-CERT, CISA-KEV) is preserved.
Within the precursor family, rank shifts with the ecosystem: MalwareBazaar
(2020) and CISA-KEV (2021) cannot appear pre-2019 and rise sharply after.
Mission character is stable; portfolio composition evolves and should be
recomputed periodically. Given the small pre-2019 subset, the claim is
made only for the precision-side cluster as a whole and the rank-one
position.

\textbf{Geographic partition.} We split by primary target jurisdiction (24
US-target attacks, 30 non-US). Vendor-Tier1 retains rank-one greedy
position in both subsets (83\% standalone US, 73\% standalone non-US).
The second source differs: US closes to 92\% via CISA-KEV and 100\% via
US-CERT; non-US closes to 93\% via US-CERT, then four precursors tie at
$+1$ (NCSC-Tier1 and CERT-UA cover INC-036; MalwareBazaar and VirusTotal
cover INC-008), the \S\ref{subsec:tiebreak} rule selects CERT-UA (97\% at
$k{=}3$), and either INC-008 choice saturates at $k{=}4$, INC-008 Norsk
Hydro being covered only by the malware repositories. CERT-UA's precursor
classification survives the split; the methodology is
jurisdiction-agnostic, individual portfolios are not.

\section{Coverage-Cost Portfolio Selection}
\label{sec:portfolio}

\subsection{Greedy frontier and exhaustive verification}

At each step, greedy adds the source that maximizes marginal new-attack
coverage over the eight defender-relevant classes (broad-coverage,
precursor, plus borderline CISA-KEV) and stops at zero marginal gain.
Table~\ref{tab:greedyfrontier} reports the Pareto frontier with
percentile-method bootstrap confidence intervals (2{,}000 incident-level
resamples; incident-level resampling preserves within-incident multi-signal
correlation).

\textbf{Tie-break rule.}
\label{subsec:tiebreak}
Where two or more candidate sources contribute identical marginal coverage
at a greedy step, the next source is selected by the following
deterministic ordering:
(i) mission profile, with broad-coverage before
precursor before disclosure-exposure; (ii) within profile, alphabetical
ascending by source class name.
This rule is applied uniformly to all
greedy frontiers reported in this paper
(Tables~\ref{tab:greedyfrontier}, \ref{tab:sector}, \ref{tab:actorcell})
and is documented in the released corpus alongside the resolved order at
each step.

\begin{table}[t]
\caption{Greedy Pareto Frontier with 95\% Percentile-Method
Bootstrap CIs (2{,}000 Incident-Level Resamples).}
\label{tab:greedyfrontier}
\centering
\footnotesize
\setlength{\tabcolsep}{4pt}
\renewcommand{\arraystretch}{1.05}
\begin{tabular}{@{}rlcc@{}}
\toprule
$k$ & \textbf{Source added} & \textbf{Cum.\ cov.\ (95\% CI)} & \textbf{Marg.\ gain}\\
\midrule
1 & Vendor-Tier1   & 77.8\% (66.7--88.9)   & +77.8 pp\\
2 & US-CERT        & 92.6\% (85.2--98.1)   & +14.8 pp\\
3 & MalwareBazaar  & 96.3\% (92.6--100.0)  & +3.7 pp\\
4 & CERT-UA        & 98.1\% (98.1--100.0)  & +1.9 pp\\
5 & CISA-KEV       & 100.0\% (100.0--100.0)& +1.9 pp\\
\bottomrule
\end{tabular}\\[2pt]
{\scriptsize Ties at $k{=}3$: MalwareBazaar and CISA-KEV both contribute 2
previously-uncovered incidents; the unified tie-break rule
(\S\ref{subsec:tiebreak}, mission-profile ordering then alphabetical)
selects MalwareBazaar (precursor) over CISA-KEV (disclosure-exposure). At
$k{=}4$, three sources tie at $+1$ incident: NCSC-Tier1 and CERT-UA both
cover INC-036, CISA-KEV covers INC-021; mission-profile ordering removes
CISA-KEV and alphabetical selects CERT-UA. Either $k{=}4$ choice
(NCSC-Tier1 or CERT-UA) yields identical 98.1\% coverage. CISA-KEV is
selected at $k{=}5$ where its remaining marginal coverage (INC-021) is
unique. Exhaustive enumeration over the eight defender-relevant classes
confirms greedy matches the optimum at every $k$.}
\end{table}

\textbf{Bootstrap and benchmark.} 
The narrow upper CI bounds reflect a ceiling effect:
24.1\%, 76.3\%, and 100\% of the
2{,}000 incident-level resamples reach 100\% coverage at $k{=}3,4,5$
respectively. Exhaustive enumeration of three-source subsets over the
seven precursor-and-broad-coverage classes (35 combinations) benchmarks
the greedy frontier
(CISA-KEV is excluded from the random-baseline
universe because its disclosure-exposure profile would inflate the
apparent value of source-aware selection)
Greedy reaches 96.3\% at
$k{=}3$ (52 of 54); random three-source mean is 56.5\% (sd 28.3 pp,
min 22.2\%, i.e.\ 12 of 54). The 39.8 pp greedy-vs-mean-random gap
quantifies the value of source-aware portfolio construction. Inter-source
signal overlap is handled directly by the set-cover semantics, computed
from per-signal incident mappings in the released corpus.
At this instance size exhaustive enumeration, not the $(1{-}1/e)$
guarantee, certifies optimality; greedy is retained for scale.

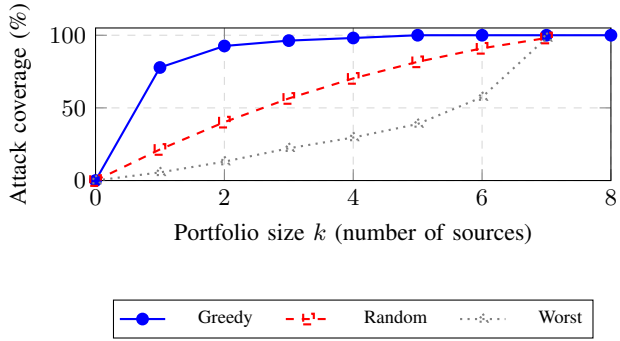
\begin{figure}[t]
\centering
\begin{tikzpicture}
\begin{axis}[
  width=8.4cm, height=3.6cm,
  xlabel={Portfolio size $k$ (number of sources)},
  ylabel={Attack coverage (\%)},
  xmin=0, xmax=8, ymin=0, ymax=105,
  grid=major, grid style={dashed,gray!30},
  legend style={at={(0.5,-0.78)}, anchor=north, font=\scriptsize,
    legend columns=3, column sep=10pt, inner sep=2pt},
  tick label style={font=\small},
  xlabel style={font=\small}, ylabel style={font=\small},
]
\addplot[thick, blue, mark=*, mark size=2pt] coordinates {
  (0,0) (1,77.8) (2,92.6) (3,96.3) (4,98.1) (5,100) (6,100) (7,100) (8,100)
};
\addlegendentry{Greedy}
\addplot[thick, dashed, red, mark=square, mark size=2pt] coordinates {
  (0,0) (1,21.4) (2,40.2) (3,56.5) (4,70.3) (5,81.7) (6,91.0) (7,98.1)
};
\addlegendentry{Random}
\addplot[thick, dotted, gray, mark=triangle, mark size=2pt] coordinates {
  (0,0) (1,5.6) (2,13.0) (3,22.2) (4,29.6) (5,38.9) (6,57.4) (7,98.1)
};
\addlegendentry{Worst}
\end{axis}
\end{tikzpicture}
\caption{Pareto frontier of CI attack coverage by portfolio size.
Greedy reaches 96.3\% at $k{=}3$ versus 56.5\% mean random and 22.2\%
worst-case across 35 three-source subsets of the seven precursor and
broad-coverage classes.}
\label{fig:pareto}
\end{figure}

\subsection{Commercial collection scope and marginal value}
\label{sec:commercial}

The Vendor-Tier1 class aggregates research output from Tier-1 industry
vendors and major incident-response firms (Mandiant, CrowdStrike, Symantec,
Kaspersky, Palo Alto Unit 42, Microsoft MSRC, Cisco PSIRT, Schneider
Electric ProductCERT, Siemens ProductCERT, Dragos, ESET, and a small number
of others) via public research blogs, security advisories, and threat
reports. Collection was performed against publicly accessible analyst output
at the same effort level as free-source collection.
We did not procure enterprise paid feeds (untested: CrowdStrike Falcon
Intelligence, Flashpoint, Group-IB, Intel 471, ZeroFox, Anomali,
Cybersixgill, Mandiant TI paid tiers).
Bouwman et
al.~\cite{bouwman2020tivalue}, with full paid-feed access to two leading
vendors, found vendor-to-vendor overlap of only 2.5 to 4.0 percent on tracked
threat actors, suggesting paid feeds add limited coverage value beyond what
we measure even at enterprise scale.

The marginal value of the Vendor-Tier1 class above a purely-free portfolio
is the relevant operational quantity. Removing Vendor-Tier1 from the
defender portfolio and re-running greedy on the remaining seven defender
classes produces the ordering US-CERT (27.8\%) $\to$ CISA-KEV (cum.\ 44.4\%)
$\to$ MalwareBazaar (51.9\%) $\to$ NCSC-Tier1 (57.4\%) $\to$ OTX (63.0\%)
$\to$ VirusTotal (68.5\%) $\to$ CERT-UA (70.4\%), saturating the eight-class
portfolio without Vendor-Tier1 at 70.4\%.
Vendor-Tier1's rank-one marginal coverage is therefore 29.6 pp, far
above the 2.5--4.0\% per-vendor overlap of~\cite{bouwman2020tivalue}.

\section{Sector, Actor-Type, and Cross-Stratified Portfolios}
\label{sec:strat}

Table~\ref{tab:sector} reports the greedy frontier recomputed by sector for
the sectors with at least three attacks. Table~\ref{tab:actorcell} reports
actor-type stratification and the four sector-by-actor cells with at least
five attacks. The cells with fewer than five attacks (eight of thirteen
sector strata, most sector-by-actor cells) are excluded from
recommendations; they remain in the released corpus for re-computation as
additional incidents accumulate.

\begin{table}[!htb]
\caption{Greedy Per-Sector Portfolios ($n\geq 3$). Coverage percentages are
within-stratum.}
\label{tab:sector}
\centering
\footnotesize
\setlength{\tabcolsep}{3pt}
\renewcommand{\arraystretch}{1.05}
\begin{tabular}{@{}lr p{0.66\columnwidth}@{}}
\toprule
\textbf{Sector} & $n$ & \textbf{Greedy portfolio}\\
\midrule
Energy        & 15 & Vendor-Tier1 (87\%) $\to$ US-CERT (100\%)\\
Healthcare    & 14 & Vendor-Tier1 (79\%) $\to$ CISA-KEV (93\%) $\to$ US-CERT (100\%)\\
Manufacturing &  5 & Vendor-Tier1 (60\%) $\to$ CERT-UA (80\%) $\to$ MalwareBazaar (100\%)\\
Transport     &  3 & US-CERT (67\%) $\to$ Vendor-Tier1 (100\%)\\
Water         &  3 & Vendor-Tier1 (67\%) $\to$ US-CERT (100\%)\\
Government    &  3 & Vendor-Tier1 (67\%) $\to$ CERT-UA (100\%)\\
\bottomrule
\end{tabular}
\end{table}

\begin{table}[!htb]
\caption{Greedy Portfolios by Actor Type and by Sector-by-Actor Cells
($n\geq 5$).}
\label{tab:actorcell}
\centering
\footnotesize
\setlength{\tabcolsep}{3pt}
\renewcommand{\arraystretch}{1.05}
\begin{tabular}{@{}lr p{0.66\columnwidth}@{}}
\toprule
\textbf{Stratum} & $n$ & \textbf{Greedy portfolio}\\
\midrule
Cybercriminal  & 34 & Vendor-Tier1 (74\%) $\to$ US-CERT (88\%) $\to$ MalwareBazaar (94\%)\\
Nation-State   & 17 & Vendor-Tier1 (94\%) $\to$ US-CERT (100\%)\\
\midrule
Healthcare/Cyber.    & 12 & Vendor-Tier1 (75\%) $\to$ CISA-KEV (92\%) $\to$ US-CERT (100\%)\\
Energy/Nation-State  &  9 & Vendor-Tier1 (89\%) $\to$ US-CERT (100\%)\\
Energy/Cyber.        &  5 & Vendor-Tier1 (100\%)\\
Manufact./Cyber.     &  5 & Vendor-Tier1 (60\%) $\to$ CERT-UA (80\%) $\to$ MalwareBazaar (100\%)\\
\bottomrule
\end{tabular}
\end{table}

The stratification yields three structural observations. First, Vendor-Tier1
is the rank-one source in every stratum with at least three attacks except
Transport, which is dominated by US-CERT.
Second, the closing source tracks actor mix: Cybercriminal-heavy strata
close through CERT-UA, MalwareBazaar, and CISA-KEV, reflecting tooling
that produces public samples and exploited-CVE disclosures; Nation-State
strata close through US-CERT under the tie-break (the shared residual,
Shamoon INC-002, is co-covered by US-CERT and VirusTotal). Third, small
cells should not be over-interpreted: Government ($n{=}3$) closing
through CERT-UA is corpus-composition artefact, not a recommendation; the
release supports recomputation as cells grow.

\section{Adaptive Adversary Analysis}
\label{sec:adversary}

\begin{table}[!htb]
\caption{Attack-Precursor Coverage Under Adversary Suppression Scenarios
(Computed Against the 54-Attack Corpus).}
\label{tab:adversary}
\centering
\footnotesize
\setlength{\tabcolsep}{3pt}
\renewcommand{\arraystretch}{1.05}
\begin{tabular}{@{}p{0.30\columnwidth} p{0.40\columnwidth} c@{}}
\toprule
\textbf{Scenario} & \textbf{Sources retained} & \textbf{Cov.}\\
\midrule
Full recommended portfolio ($k{=}5$)
   & VT1 + US-CERT + MB + CERT-UA + KEV         & 100.0\%\\
Vendor-Tier1 suppressed
   & 7 non-Vendor sources                       &  70.4\%\\
Malware repos only suppressed (Vendor retained)
   & VT1 + US-CERT + NCSC + CERT-UA + KEV       &  98.1\%\\
Vendor, all repos, and KEV suppressed (gov.-CSIRT only)
   & US-CERT + NCSC + CERT-UA                   &  38.9\%\\
\bottomrule
\end{tabular}
\end{table}

The recommended-portfolio sources differ in structural suppressibility.
\emph{US-CERT} and \emph{NVD/NIST} follow confirmed victim impact and CVE
disclosure respectively, neither adversary-controllable. \emph{Vendor-Tier1}
operates predominantly at the TTP and infrastructure-attribution level,
populated through forensic work the adversary cannot suppress.
\emph{MalwareBazaar} and \emph{VirusTotal} are partially suppressible
through sample-handling discipline; the 5 MalwareBazaar and 7 VirusTotal
verified signals in the corpus were submitted by entities other than the
attackers in every recorded case. \emph{NCSC-Tier1} requires coordinated
disruption of multiple sovereign government channels to suppress;
\emph{CERT-UA} reflects national CSIRT operations under conflict
conditions and is difficult to suppress at the source.

\textbf{Coverage degradation under suppression.} Table~\ref{tab:adversary}
quantifies portfolio degradation across four scenarios.


A defender losing Vendor-Tier1 entirely retains 70.4\% attack-precursor
coverage from sources the adversary cannot structurally suppress. Reduced
to purely government-CSIRT channels (no vendor research, no
malware-repository or community-exchange sampling, no CISA-KEV disclosure),
coverage drops to 38.9\%. The gap identifies operational exposure to
adversary suppression discipline; periodic recomputation of empirical
priors against fresh incident data is the recommended mitigation for
landscape drift.

\section{Walked Examples and Lead-Time Analysis}
\label{sec:walked}

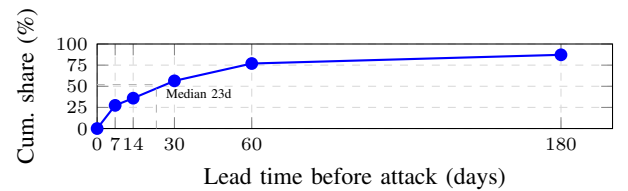
\begin{figure}[!b]
\centering
\begin{tikzpicture}
\begin{axis}[
  width=8.4cm, height=2.7cm,
  xlabel={Lead time before attack (days)},
  ylabel={Cum.\ share (\%)},
  xmin=0, xmax=200, ymin=0, ymax=100,
  grid=major, grid style={dashed,gray!30},
  tick label style={font=\scriptsize},
  xlabel style={font=\small}, ylabel style={font=\small},
  ytick={0,25,50,75,100}, xtick={0,7,14,30,60,180},
]
\addplot[thick, blue, mark=*, mark size=2pt] coordinates {
  (0,0) (7,27.4) (14,35.9) (30,56.4) (60,76.9) (180,87.2)
};
\draw[dashed, gray!60] (axis cs:23,0) -- (axis cs:23,52.0) -- (axis cs:0,52.0);
\node[font=\tiny, anchor=north west, fill=white, inner sep=0.6pt] at (axis cs:26, 49) {Median 23d};
\end{axis}
\end{tikzpicture}
\caption{Empirical CDF of attack-precursor signal lead times across 117
verified attack-case occurrences.}
\label{fig:leadtime}
\end{figure}

\textbf{Colonial Pipeline (May 7, 2021)} appears in the verified corpus as
INC-009 with two pre-attack Vendor-Tier1 signals: a report on the
emergence of the DarkSide RaaS group 270 days before impact, and a report
on DarkSide affiliate recruitment and TTPs 16 days before.
CISA AA21-131A~\cite{cisaaa21131a}, published four days after the attack,
is recorded under CISA-KEV at $\mathrm{dba}{=}{-}4$: it counts toward the
class's attack coverage but not toward the lead-time analysis of
Figure~\ref{fig:leadtime}.

\textbf{MOVEit Transfer (May 27, 2023)} appears as INC-020 and is the
operational-near-zero-day limit case the inclusion criterion was designed
to admit. 
The qualifying signal row (SIG-0069) is recorded under the NVD class at
$\mathrm{dba}{=}0$ using the first-exploitation date, a convention
documented in the released row; the NVD record itself published June 2
($\mathrm{dba}{=}{-}6$), the Progress Software advisory followed at
$\mathrm{dba}{=}{-}4$, and CISA AA23-158A~\cite{cisaaa23158a} at
$\mathrm{dba}{=}{-}11$, all post-attack, so INC-020's inclusion rests on
the same-day convention rather than a strictly pre-attack publication.
A Cl0p dark-web disclosure at $\mathrm{dba}{=}{-}19$ was tracked but
excluded on linkage-protocol grounds. MOVEit thus reports zero positive
lead days in v3.1, with the same-day NVD entry qualifying; such
near-zero-day cases are flagged per incident and bound the framework's
lead-time ceiling, not its coverage ceiling.

\textbf{Change Healthcare ALPHV/BlackCat (February 21, 2024)} appears as
INC-021 with four verified pre-attack signals: the CISA-KEV
AA23-353A~\cite{cisaaa23353a} joint advisory on ALPHV/BlackCat targeting
healthcare at 64 days,
two NVD entries for CVE-2024-1709 ConnectWise
ScreenConnect authentication bypass at 2 days, and a public GitHub PoC for
the same CVE on day zero.
Only AA23-353A enters the eight-class defender calculation (the NVD rows
and the GitHub PoC are out of class). The top-two portfolio misses
INC-021; canonical rank 3 selects MalwareBazaar under the tie-break and
still misses it, while the co-tied alternate CISA-KEV recovers it through
the
64-day
advisory. It is the corpus's cleanest exposure of the
precursor-versus-disclosure tradeoff at $k{=}3$; canonical greedy reaches
CISA-KEV at $k{=}5$, and the Healthcare/Cybercriminal portfolio of
Table~\ref{tab:actorcell} recommends CISA-KEV at rank 2 for this reason.

\textbf{Volt Typhoon} (INC-018, anchor February 7, 2024; multi-year
pre-positioning campaign anchored on AA24-038A~\cite{cisaaa24038a}) is the
clearest specialized-precursor demonstration in the corpus.
Two pre-attack
signals appear at 259 days: AA23-144A~\cite{cisaaa23144a} (PRC
State-Sponsored Living off the Land, May 24, 2023), and a Vendor-Tier1
publication (Microsoft Threat Intelligence on Volt Typhoon, same date).
AA23-144A is recorded under CISA-KEV; early capture is dominated by
Vendor-Tier1, the case where vendor research outruns pure government
channels.

\textbf{Lead-time distribution.} Across 117 attack-case verified-signal
occurrences dated at or before the verified attack date (97 strictly
positive plus 20 same-day), 27.4\% appear within 7 days, 35.9\% within 14,
56.4\% within 30, 76.9\% within 60, and 87.2\% within 180 (median 23 days,
IQR 6--59). Figure~\ref{fig:leadtime} plots the empirical CDF.
The 23-day median implies weekly review discards about half the available
warning; daily computation recovers most of it. The 259-plus-day tail
belongs to specialized nation-state pre-positioning cases.


\section{Discussion}
\label{sec:discussion}

The mission separation is the central contribution: the pooled split is
highly significant ($p = 3.4 \times 10^{-8}$) and the three-profile
structure is stable across both partitions and nearly every stratum.
\emph{Portfolio sizing}: two
precursor-plus-broad-coverage sources cover 92.6\% of corpus attacks,
three cover 96.3\%; the 23-day median lead time means daily portfolio
computation recovers warning that weekly review discards.
\emph{Stratum specificity}: portfolios share Vendor-Tier1 as rank-one
anchor but diverge at ranks two and three by sector and actor mix
(Healthcare/Cybercriminal closes through CISA-KEV;
Manufacturing/Cybercriminal through CERT-UA and MalwareBazaar;
Energy/Nation-State through US-CERT). 
\emph{Queue separation}:
disclosure-exposure sources belong to a patch-management queue, not
incident response; conflation is the largest preventable analyst-fatigue
source.
\emph{Network-defense integration}: repository IOCs feed IDS/EDR rules,
NVD feeds scanners and patch prioritization, Joint Advisory IOCs feed
denylists. The framework generalizes to any threat-intelligence consumer
with bounded attention and an attack history.
That includes operational-technology fleets in adjacent domains such as
robotics, which consume the same advisory and vulnerability channels
measured here.

\section{Limitations and Threats to Validity}
\label{sec:limits}

\textbf{L1. Corpus size and quality residuals.} 54 attacks and 12 nulls limit
per-source null-side power; three classes survive BH-FDR at 0.05
(Vendor-Tier1, NVD, ICS-CERT) while six precursor classes are point-clean
but underpowered. The pooled profile-family separation is highly significant
($p \approx 3.4 \times 10^{-8}$); bootstrap CIs at $k{=}1{-}3$ are $\pm 10$
to 15 pp.
Two release residuals remain with recorded sign-off: gate G6 (two-signal
floor) is violated for INC-030/035/041/043, and INC-021 carries duplicate
NVD rows (SIG-0074, SIG-0184) for CVE-2024-1709, harmless to precision
but inflating per-incident NVD density.

\textbf{L2. Construction-bounded coverage.} Inclusion required at least one
observable pre-attack signal, biasing toward attacks with public footprint.
Silent attacks contribute no observable signals, so source classification
is invariant under uniform silent-attack scaling; absolute coverage figures
degrade proportionally.

\textbf{L3. Commercial-product sample.} Publicly accessible vendor research
was collected at the same effort as free-source collection; no paid feeds
were procured. Bouwman et al.~\cite{bouwman2020tivalue} report paid feeds
add limited coverage at enterprise scale.

\textbf{L4. Publication versus consumption.} We measure whether a source
published a signal before an attack, not whether a defender observed and
acted on it. Results bound theoretical visibility from public OSINT, not
operational effectiveness.
For the same reason, within-corpus precision does not model per-day alert
volume (false-alarm load against live streams is future work), and
coverage percentages are conditional on an attack being OSINT-visible
enough to enter the corpus under the \S\ref{sec:audit} criteria; they are
not detection rates over all CI attacks.

\textbf{L5. Temporal scope.} The fifteen-calendar-year span (May
2010--October 2024) covers significant landscape changes (MalwareBazaar
2020; CISA KEV 2021). The temporal partition (\S\ref{sec:predval}) shows
precision-side classification is stable across the 2019 regime change while
source rank drifts; periodic recomputation is required. Formal adversary
modeling is open.
External validation against a post-2024 holdout or a live SOC deployment
also remains open; the \S\ref{sec:predval} partitions are within-corpus
stability checks.

\textbf{L6. Annotation protocol.} See \S\ref{sec:audit}: 84.5\%
agreement, 12.0\% conflict envelope.

\textbf{Sensitivity recomputes.} Headline results are robust to two
restrictive recomputes.
(i) Strict pre-attack only ($\mathrm{dba}\geq 0$): coverage shifts
modestly (VT1 77.8$\to$74.1, US-CERT 27.8$\to$25.9, VirusTotal
11.1$\to$9.3, CISA-KEV 16.7$\to$11.1, ICS-CERT 7.4$\to$5.6\%; others
unchanged); precursor null firings stay zero; the pooled table becomes
$[37,0;20,23]$, $p=1.4\times 10^{-8}$; greedy saturates at 96.3\% because
INC-001 (Stuxnet) and INC-020 (MOVEit) are NVD-only at
$\mathrm{dba}\geq 0$.
(ii) Strict-linkage only (criteria (a)--(d), 129 of 161 signals):
precursor classes keep zero null firings, disclosure-exposure classes keep
null contamination at or above coverage, Vendor-Tier1 keeps rank one
(66.7\% standalone), and the pooled separation survives ($[17,0;24,22]$,
one-sided Fisher exact $p=1.5\times 10^{-4}$); government-advisory
precursor coverage (US-CERT, NCSC-Tier1, OTX) falls sharply because its
linkage runs through (e)--(g) while the malware-repository classes are
unchanged, so coverage magnitudes are conditional on boundary linkage and
the precision-side taxonomy is not.

\textbf{Anticipated critiques.} \emph{EPSS:} EPSS~\cite{jacobs2021epss}
operates per-CVE; we operate per-source. Defenders use the recommended
portfolio for source-level monitoring and EPSS for per-CVE prioritization
within those signals.
\emph{Vendor-Tier1 dominance and CISA-KEV aggregation:} vendor writeups
are where most public attribution and IOC disclosure occurs; per-vendor
decomposition lowers any single class while preserving combined
portfolios. CISA-KEV aggregates the KEV catalog (firmly
disclosure-exposure) with Joint Advisories (closer to precursor),
preserved from the v2 registry and decomposable from the release; the
borderline aggregate behavior is itself informative.
\emph{External validation:}
the full corpus (incidents, signal records, linkage decisions) and the
classification rules of \S\ref{sec:taxonomy} are released for independent
re-derivation.

\section{Conclusion}
\label{sec:conclusion}

We provide an evidence-based answer to which OSINT sources caught critical
infrastructure cyberattacks across a fifteen-calendar-year null-controlled
corpus of 54 confirmed attacks and 12 nulls. Public OSINT source classes
separate empirically into three operational profiles (pooled Fisher exact
$p = 3.4 \times 10^{-8}$): precursor (six classes, zero null firings),
disclosure-exposure (three classes whose null contamination meets or
exceeds attack coverage), and one broad-coverage class with 77.8\% standalone
attack coverage and 91.3\% within-corpus precision. The precision-side
classification is stable across a 2019 temporal partition and a
US-versus-non-US geographic partition.
The taxonomy's departures from industry conventional wisdom for several
heavily-promoted sources are structural, not evaluative. Two sources cover 92.6\% of corpus attacks, three 96.3\%,
and greedy at $k{=}3$ beats random by 39.8 pp; portfolios diverge by
stratum (Healthcare/Cybercriminal via CISA-KEV,
Manufacturing/Cybercriminal via CSIRT and repositories,
Energy/Nation-State via US-CERT).
The corpus, linkage protocol, and
classification rules are released at
\url{https://github.com/jeelkhatiwala/IEEE-CNS-Dataset}.

\balance


\section*{Ethical Considerations}
All data are drawn from publicly accessible OSINT; no privileged, hacked,
or non-public sources were used. Malware samples referenced for linkage
were handled in isolated environments; no live payloads are distributed.
Characterizing source effectiveness could inform adversary suppression,
but the taxonomy formalizes patterns already legible to sophisticated
actors and primarily benefits under-resourced defenders;
\S\ref{sec:adversary} quantifies coverage retained against
non-suppressible sources.

\section*{Reproducibility and Artifact Availability}
The released artifact is the corpus dataset: incident records with
Tier-1 verification URLs, signal records with linkage decisions and
lead-day values, the null-selection log, and the round-2
release-preparation worksheet (51 rows). Tables and figures reproduce
from the CSVs with standard statistical software; no project code is
required.

\bibliographystyle{IEEEtran}

\end{document}